\documentclass[prl,twocolumn,floatfix,amsmath,nofootinbib,amssymb,showpacs]{revtex4}

\usepackage{graphicx}
\usepackage{dcolumn}
\usepackage{mathrsfs}
\usepackage{bm}
\usepackage{color,xspace}
\usepackage[small,compact,raggedright]{titlesec}
\usepackage{epstopdf}

\newcommand{\diff}{\text{d}}

\begin{document}

\title{Magnetic Charge Can Locally Stabilize Kaluza-Klein Bubbles}
 \author{Sean Stotyn}
 \email{smastoty@sciborg.uwaterloo.ca}
 \author{Robert B. Mann}
 \email{rbmann@sciborg.uwaterloo.ca}
\affiliation{Department of Physics \& Astronomy, University of Waterloo, \\Waterloo, Ontario Canada N2L 3G1}

\begin{abstract}

We construct a new 2-parameter family of static topological solitons in 5D minimal supergravity which are endowed with magnetic charge and mass.  The solitons are asymptotically ${\mathbb R}^4\times S^1$, where the radius of the $S^1$ has a lower bound $R_s\ge R_{min}$.  Setting up initial data on a Cauchy slice at a moment of time symmetry, we demonstrate that if $R_s>R_{min}$ these solitons correspond to a perturbatively stable ``small" static bubble as well as an unstable ``large" static bubble, whereas if $R_s<R_{min}$ there are no static bubbles.  The energetics and thermodynamics of the magnetic black string are then discussed and it is shown that the locally stable bubble is the end point of a phase transition for an appropriate range of black string parameters.

\end{abstract}

\date{2011}

\pacs{04.50.Cd, 04.50.Gh, 04.70.Dy, 04.65.+e}


\maketitle

Since the seminal papers of Hawking and Page \cite{Hawking:1982dh} and of Gregory and Laflamme \cite{Gregory:1993vy}, the stability of black objects, particularly in various higher dimensional gravity theories, has become a particularly active vein of research.  In the modern parlance of the AdS/CFT correspondence, the Hawking-Page phase transition of AdS black holes, which is thermodynamic in nature, is dual to a confinement-deconfinement phase transition in the large $N$ limit of the boundary gauge theory \cite{Witten:1998wn}.  Meanwhile the qualitatively unrelated Gregory-Laflamme instability, which manifests as unstable modes in metric perturbations at the horizon, has been shown to be similar to the Rayleigh-Plateau instability of classical membranes \cite{Cardoso:2006ks}.  There are then two important types of stability when it comes to black holes: thermodynamic and perturbative.  In \cite{Gubser:2000mm} it was conjectured that there exists a relation between thermodynamic stability and perturbative stability, although this conjecture has since been shown to be violated in some cases \cite{Friess:2005zp}.

Perturbative stability, \`a la Gregory-Laflamme, of black strings with magnetic charge and mass parameters $P$ and $m$, consistent with the notation in this letter, was demonstrated to hold in the range $P\le m\le\frac3{2\sqrt{2}}P$, corresponding to where the heat capacity is positive ~\cite{Miyamoto:2007mh}.   While this was regarded as evidence that the Gubser-Mitra conjecture is satisfied for these objects, a proper thermodynamic stability analysis was not possible since it was unclear what the magnetic black string could phase transition to.  For neutral and electrically charged black strings, there exist spherical black holes that are (un)charged under the same gauge field as the string.  This is not true for strings carrying a topological magnetic charge since this charge must have flux on a sphere of the same dimension as the rank of the gauge field.   This means that spherical magnetically charged black holes are charged under a larger rank gauge field than the black string, and so charge conservation must be violated to allow a transition between these objects.  The question is then begged: do there exist other objects with the same topological charge as the black string?  Here we demonstrate that if one considers anti-periodic boundary conditions for the fermions, there indeed exists a 2 parameter family of solitons that carry the same topological charge as the black string. This 2-parameter family is sub-divided into ``small" and ``large" solitons, which we will show contribute dramatically to the thermodynamic stability of magnetic black strings.  

 It is, however, well known that Kaluza-Klein spacetimes with supersymmetry breaking boundary conditions have rather fatal instabilities.  For instance, in \cite{Witten:1981gj} Witten showed that in empty space with a Kaluza-Klein direction, ``bubbles of nothing" can nucleate and expand exponentially fast, eating up the entire spacetime.  Brill and Horowitz \cite{Brill:1991qe} then expanded on the work of Witten by considering an asymptotically Kaluza-Klein spacetime with initial data on a Cauchy surface of time symmetry and demonstrating that such spacetimes with anti-periodic fermionic boundary conditions are unstable to the nucleation of dynamic bubbles of arbitrary negative energy.  These results hold for vacuum Einstein as well as Einstein-Maxwell with electric charge and magnetic charge with flux around the periodic dimension.  Further investigations have revealed that topological solitons carrying the appropriate Maxwell charges in these theories generically correspond to perturbatively unstable static bubbles \cite{Sarbach:2004rm,Sarbach:2003dz,Dine:2006we} and are therefore a subset of the bubbles considered in \cite{Brill:1991qe}. 

Bubbles carrying a topological magnetic charge are not covered by the above analysis.  It is our aim to demonstrate  that the addition of a topological magnetic charge locally stabilizes the Kaluza Klein vaccum\footnote{It should be understood that this is a local vacuum.  There are bubbles with arbitrary negative energy and so a true vacuum cannot be defined.}, where the vacuum is a small static magnetically charged topological soliton.

We start with the metric of the static, magnetically charged black string of 5D minimal supergravity \cite{Miyamoto:2007mh}
\begin{eqnarray}
&&ds^2=-f(r)dt^2+\frac{dr^2}{f(r)h(r)}+h(r)dz^2+r^2d\Omega_2^2 \label{eq:Metric}\\
&&{\bf A}=-\sqrt{3}P\cos\theta \diff\phi,\\
&&f(r)=1-\frac{r_+}r, \>\>\>\>\>\>\>\>\>\>\>\>\>\>\>\>\>\> h(r)=1-\frac{r_-}r,
\end{eqnarray}
where $d\Omega_2$ is the metric on a unit two-sphere, $\bf A$ is the gauge potential and the outer and inner horizons are $r_\pm=m\pm\sqrt{m^2-P^2}$.  The magnetic charge and the ADM mass of the black string are related to the parameters $m$ and $P$ as
\begin{equation}
Q_M=\sqrt{3}P,\qquad 
M_{BS}=\frac{\pi R_s}{2G_5}\left(3m+\sqrt{m^2-P^2}\right) \label{eq:BSMass}
\end{equation}
where $R_s$ corresponds to the radius of the compact $z$ direction and $G_5$ is the five dimensional Newton's constant.  The extremal solution is given by $m^2=P^2$ and takes the explicit form
\begin{equation}
ds^2=\left(1-\frac{P}{r}\right)(-dt^2+dz^2)+\frac{dr^2}{\left(1-\frac{P}{r}\right)^2}+r^2d\Omega^2, \label{eq:MetricExtremal}
\end{equation}
which is the extremal $M5^3$ string of \cite{Kim:2010bf} if anti-periodic fermionic boundary conditions are taken.

We observe that $\partial_t$ and $\partial_z$ are Killing directions and that the gauge field has the form ${\bf A}=A_\phi {\mathrm d}\phi$, which permits us to Wick rotate (\ref{eq:Metric}) as follows: $t\rightarrow iz$ and $z\rightarrow it$.  This operation yields the metric of a static magnetically charged soliton:
\begin{eqnarray}
&&ds^2=-\tilde f(r)dt^2+\frac{dr^2}{\tilde f(r)\tilde h(r)}+\tilde h(r)dz^2+r^2d\Omega^2 \label{eq:MetricSol}\\
&&{\bf A}=-\sqrt{3}P \cos\theta\diff\phi\\
&&\tilde f(r)=1-\frac{r_c}r, \>\>\>\>\>\>\>\>\>\>\>\>\>\>\>\>\>\> \tilde h(r)=1-\frac{r_s}r,
\end{eqnarray}
where the critical radii are now $r_s=\mu+\sqrt{\mu^2-P^2}$ and $r_c=\mu-\sqrt{\mu^2-P^2}$.  
The magnetic charge and the ADM mass of the soliton are 
\begin{equation}
Q_M=\sqrt{3}P,\qquad
M_{Sol}=\frac{\pi R_s}{2G_5}\left(3\mu-\sqrt{\mu^2-P^2}\right). \label{eq:SolitonMass}
\end{equation}

The signature of the metric in the region $r_c<r<r_s$ is $(---++)$, so the solitons are static bubbles of radius $r_s$.  To ensure that there is no conical singularity at $r=r_s$, we require $z=R_s\sigma$ such that $\sigma$ has period $2\pi$ and
\begin{equation}
R_s=2 \sqrt{\frac{r_s^3}{r_s-r_c}}.\label{eq:Rs}
\end{equation}
Rearranging for $r_s$ we find $r_s^2=\frac{R_s^2}{8}\left(1\pm\sqrt{1-\frac{16P^2}{R_s^2}}\right)$, meaning that a given compactification radius admits a large soliton ($+$ sign) and a small soliton ($-$ sign) and furthermore these two solitons degenerate at a minimum compactification radius of $R_s=4P$.  These solitons also admit an ``extremal" limit obtained by taking $\mu^2=P^2$.  Such an extremal limit again yields the metric of Eq.~(\ref{eq:MetricExtremal}), as could be anticipated since this metric is invariant under Wick rotation of $t$ and $z$.

Having constructed the static bubble solutions, we turn now to constructing general bubble solutions that are dynamic.  We choose a Cauchy surface, $\Sigma$, at a moment of time symmetry so that the extrinsic curvature vanishes:
\begin{equation}
ds_{\Sigma}^2=\frac{dr^2}{f(r)h(r)}+r^2d\Omega^2+h(r)R_s^2d\sigma^2 \label{eq:InitialData}
\end{equation}
where, to ensure our static bubbles are contained in the initial data, we choose $f(r)=1-\frac{r_c}r$ and $h(r)$ is determined by the Hamiltonian constraint:
\begin{equation}
^4{\cal R}=\frac14F_{ab}F^{ab}.
\end{equation}
Here $^4{\cal R}$ is the Ricci scalar of the Cauchy surface and $F_{ab}F^{ab}=\frac{6P^2}{r^4}$ is the square of the gauge field projected onto the Cauchy surface.  With our choice of $f(r)$, the Ricci scalar is
\begin{equation}
^4{\cal R}=-\frac{2r(r-r_c)h''+(8r-7r_c)h'+4(h-1)}{2r^2}
\end{equation}
where a prime denotes a derivative with respect to $r$.  The function $h(r)$ satisfying the Hamiltonian constraint is
\begin{eqnarray}
&&h(r)=1-\frac{r_s}{r}+\frac{\sqrt{f(r)}}{r^2}C_1\label{eq:h}\\
&&+\frac{2(r-3r_c)+3\sqrt{f(r)}r_c\ln\left(r-\frac{r_c}{2}+r\sqrt{f(r)}\right)}{r^2}C_2\nonumber
\end{eqnarray}
where $C_1$ and $C_2$ are constants of integration  and we have used $r_sr_c=P^2$ in the second term.  

Next we require $h(r)$ to have a simple zero at $r_s$, meaning $C_1$ is determined in terms of the other variables as
\begin{equation}
C_1=\frac{-2(r_s-3r_c)-3\sqrt{f_s}r_c\ln\left(r_s-\frac{r_c}{2}+r_s\sqrt{f_s}\right)}{\sqrt{f_s}}C_2 \label{eq:C1}
\end{equation}
where $f_s\equiv f(r_s)$.  Near $r=r_s > r_c$ the function $h(r)=h'(r_s)(r-r_s)+...$ where
\begin{equation}
h'(r_s)=\frac{r_s-r_c+2C_2}{r_s(r_s-r_c)}.
\end{equation}
In order for the boundary $r=r_s$ to be free of conical singularities, $R_s$ is constrained via
\begin{equation}
h'(r_s)\sqrt{f(r_s)}=\frac{2}{R_s}
\end{equation}
which can be solved for $C_2$:
\begin{equation}
C_2=\frac12\sqrt{r_s^2-P^2}\left(\frac{2r_s}{R_s}-\frac{\sqrt{r_s^2-P^2}}{r_s}\right).
\end{equation}
From Eqs.~(\ref{eq:h}) and (\ref{eq:C1}) it is clear that when $C_2=0$ the static solution (\ref{eq:MetricSol}) is recovered.  This occurs when $r_s^2=P^2$, which is the extremal string, or when $R_s=2\sqrt{\frac{r_s^3}{r_s-r_c}}$ which is the relation (\ref{eq:Rs}).  When $R_s<4P$, $C_2$ has no zeroes, meaning there are no static bubbles.

At the moment of time symmetry, the time derivative of the metric vanishes, meaning that $\partial_t$ is approximately Killing and the 5 dimensional metric momentarily takes the form of (\ref{eq:MetricSol}) with $\tilde f(r)=1-\frac{r_c}{r}$ and $\tilde h(r)$ given by Eq.~(\ref{eq:h}).  This allows us to construct an ADM energy sufficiently close to the moment of time symmetry. The physical process is that of a static black string with a well-defined ADM energy tunnelling to a bubble configuration by releasing energy in the form of radiation; the left-over energy will be the ADM energy of the bubble.  Away from the moment of nucleation, the bubble will generally expand or contract and so, at least close to the time of nucleation, the ADM energy will act as a potential the bubble ``moves" through.  The asymptotic form of the metric functions are $f(r)=1-\frac{r_c}{r}$ and $h(r)\approx1-\frac{r_s-2C_2}{r}$, which yields
\begin{equation}
M_5=\frac{2\pi R_s}{4G_5}\left(\frac{P^2}{r_s}+2r_s-\frac{2r_s}{R_s}\sqrt{r_s^2-P^2}\right). \label{eq:BubbleMass}
\end{equation}
Note that if $R_s=2\sqrt{\frac{r_s^3}{r_s-r_c}}$, corresponding to the static soliton $C_2=0$, this mass agrees with (\ref{eq:SolitonMass}), and furthermore, if $r_s=P$, corresponding to the extremal string, this mass agrees with the extremal limit of (\ref{eq:BSMass}).

\begin{figure}
\centering
\includegraphics[width=3.4 in]{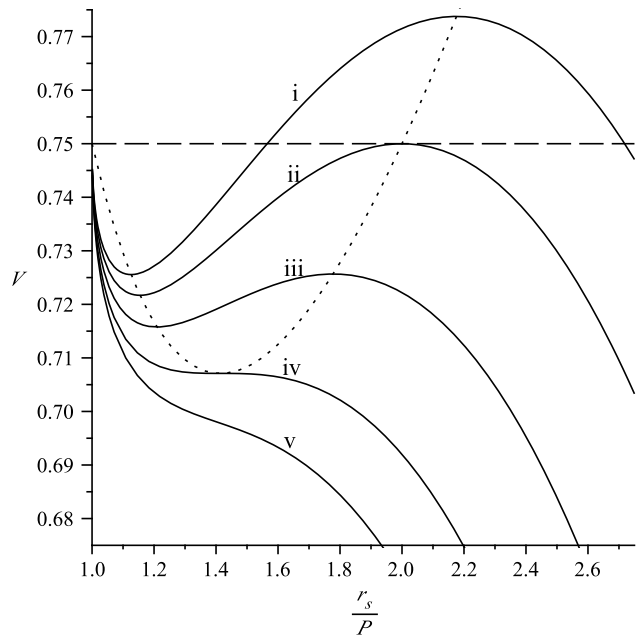}
\caption{The solid lines are the potential $V(r_s)$ for (i) $R_s=4.9P$, (ii) $R_s=\frac{8P}{\sqrt{3}}$, (iii) $R_s=4.3P$, (iv) $R_s=4P$ and (v) $R_s=3.8P$.  The dotted line shows the energy of the static solitons: the intersection of the dotted line with a potential curve occurs at an extremum of the potential.  The dashed line shows the energy of the extremal black string.  \label{fig:Potential}}
\end{figure}

In Fig.~\ref{fig:Potential} we plot the potential $V(r_s)\equiv\frac{M_5G_5}{2\pi R_s P}$ for different values of $\frac{R_s}{P}$ (solid lines) as well as plot the position of $r_s$ for large and small static solitons (dotted line) and the energy of the extremal black string (dashed line).  Wherever the dotted line crosses a solid curve, the potential has an extremum corresponding to a static soliton.  The minima represent the small solitons, the maxima represent the large solitons and the inflection point on line (iv) corresponds to where the small and large solitons degenerate at $R_s=4P$.  Line (v) has $R_s<4P$ and hence has no static solutions.  

From this potential diagram, we can see that the large solitons are perturbatively unstable while the small solitons are perturbatively stable.  In Ref.~\cite{Miyamoto:2007mh}, it was found that the magnetic black string is perturbatively stable if the mass parameter lies in the range $P\le m\le\frac3{2\sqrt{2}}P$.  If we Wick rotate $t\rightarrow iz$ and $z\rightarrow it$ to get the corresponding soliton, this range corresponds to $P\le \mu\le\frac3{2\sqrt{2}}P$ which is the parameter range of $\mu$ defining small solitons.  $\mu=\frac3{2\sqrt{2}}P$ corresponds to the minimum compactification radius, $R_{min}=4P$, where large and small solitons degenerate.  We then see that the perturbative stability of the solitons is directly related to the perturbative stability of the black string via a Wick rotation.  We further note that although the exact evolution of the dynamic bubbles is not known, the small static bubble will remain perturbatively stable and the large one unstable.  This is because although the potential may evolve in time, the positions of the minima and maxima in Fig.~\ref{fig:Potential} are time independent.

Recall that in setting up the initial data on the Cauchy slice, we chose a specific form for $f(r)$ and determined $h(r)$ via the Hamiltonian constraint.  Such initial data was found to contain our static soliton solutions, one of which is locally stable within this phase space slice.  To verify that the small soliton is truly stable, we choose an orthogonal slice through phase space by fixing $h(r)=1-\frac{r_s}{r}$ and now determine $f(r)$ via the Hamiltonian constraint.  It is conceivable that in this orthogonal slice, the small soliton might be unstable.  Omitting the details,  it is not difficult to show that for this choice of initial data with $f(r)$ vanishing at some radius $r_c<r_s$ such that $r_cr_s=P^2$, there is a unique solution given by $f(r)=1-\frac{r_c}{r}$.  This initial data yields only the static solution so there is no orthogonal direction in phase space to which the small soliton could be unstable.  We conclude, then, that the small soliton is indeed locally stable.

In a thermodynamic analysis, it is not the energy but rather the \emph{free} energy that is of importance in determining the phase structure.  To this end, we consider the Euclidean path integral approach to semiclassical gravity to compute the actions of the bubbles and the black string.  The action used is 5D Einstein-Maxwell with a Gibbons-Hawking counterterm and a Mann-Marolf counterterm \cite{Mann:2006bd,Mann:2005yr}, which renders the actions finite and does not rely on ambiguous background subtractions.  Both instantons have a metric $ds^2=f(r)d\tau^2+ds_{\Sigma}^2$ where $ds_{\Sigma}^2$ is the initial data (\ref{eq:InitialData}).  The black string is given by $f(r)=1-\frac{r_+}r$, $h(r)=1-\frac{r_-}r$ while the bubbles are given by $f(r)=1-\frac{r_c}{r}$ and $h(r)$ as in Eq. (\ref{eq:h}).  The actions are found to be
\begin{eqnarray}
&&{\cal I}_b=\frac{\beta\pi R_s}{2G_5}\left(\frac{2P^2}{r_+}+r_+\right)\\
&&{\cal I}_s=\frac{\beta\pi R_s}{2G_5}\left(\frac{2P^2}{r_s}+r_s-2C_2\right)
\end{eqnarray} 
where $\beta=\frac{4\pi r_+^2} {\sqrt{r_+^2-P^2}}$ is the inverse temperature and ensures the black string instanton is free of conical singularities.  Subscripts $b$ and $s$ denote the black string and soliton bubble respectively.  

The free energies are now defined as $F_i=\beta^{-1}{\cal I}_i$.  Both solutions are put at the same temperature and have the same radius of the compact dimension so as to allow a transition that is due entirely to local phenomena at the horizon.  In keeping the transition local, we consider a black string with horizon radius $r_+$ and a bubble with radius $r_s=r_+$.  The relative free energy per unit length, ${\cal F}\equiv \frac{F_b-F_s}{2\pi R_s}$ determines the phase structure: if ${\cal F}>0$ then the bubble is preferred, if ${\cal F}<0$ then the black string is preferred and if ${\cal F}=0$ both are equally likely.  A simple calculation yields
\begin{equation}
{\cal F}=\frac{C_2}{2G_5}.
\end{equation}
meaning that the black string is thermodynamically favoured to a bubble of the same size as long as $C_2<0$.  We find generically that the condition $C_2<0$ occurs for the range of parameters lying between the two static solitons; alternatively this corresponds to the region in which the nucleated bubble would be initially contracting (see Fig.~\ref{fig:Potential}).  We therefore conclude that the black string of horizon radius $r_+$ will tunnel to a bubble of equal size as long as the nucleated bubble is initially expanding, otherwise it will be stable.

The extremal black string, corresponding to $r_s=P$ in Fig.~\ref{fig:Potential}, is always unstable to decay to a bubble and has the lowest energy in the family of black strings, so a natural question to ask is whether such an extremal string is stable against forming a bubble that will expand out to consume the entire spacetime.  From Fig.~\ref{fig:Potential} it is clear that lines (iii), (iv) and (v) will allow this since the bubble will form arbitrarily close to $r_s=P$, giving it enough energy to continue down the potential indefinitely.  However, line (ii) has the peak of the hump at the same energy as the extremal string so as long as $R_s>\frac8{\sqrt{3}}P$, as in line (i), the extremal string is stable, at least classically.  In this case the nucleated bubble will oscillate between the two turning points, settling down to the static small soliton by emitting gravitational radiation.

Strictly speaking, we should worry about quantum tunneling through the potential barrier.  However, if the compactification radius is taken to be large compared to the magnetic charge, the supersymmetry breaking boundary conditions are arbitrarily weak.   In this case, the height and thickness of the potential barrier become arbitrarily large so that tunnelling through the barrier can be ignored and the small soliton effectively becomes the vacuum, provided the black string's mass is sufficiently small compared to $R_s$.  The black string will then Hawking evaporate until the horizon radius approaches the size of the small static soliton, at which point it will have a non-zero probability of tunneling to the local vacuum.

Apart from anti-periodic boundary conditions for the fermions, no stringy considerations have been required in our analysis up to this point. Such considerations are, however, quite relevant to the physics of 
 Kaluza Klein bubbles.  For instance, when the size of the periodic dimension shrinks below the string scale, strings wrapped around this cycle become tachyonic\cite{Rohm:1983aq}.  Furthermore,  when this tachyon condensation is localised, it induces a topology changing transition \cite{Adams:2005rb}.  

For the black strings we consider, the size of the periodic dimension at arbitrary radius is controlled by the value of $h(r)$, from which we immediately see that the circle shrinks to zero at the inner horizon (recall $h(r_-)=0$).  One must also in principle be careful about $\alpha'$ corrections. However the curvature at the outer horizon goes like $1/r_+^2$ which will remain much smaller than the string scale, even at extremality, provided $P^2\gg \ell_s^2$.  Restricting our attention to large magnetic charges, what we require in order to nucleate a bubble is for tachyon condensation to take place just outside the horizon.  Using the relation $r_-r_+=P^2$, this condition becomes
\begin{equation}
\frac{\ell_s^2}{R_s^2}=\left(1-\frac{P^2}{r_+^2}\right).
\end{equation}
In order to avoid tachyon condensation at large $r$, we also require $R_s^2\gg\ell_s^2$, which implies that the black string is near extremality.

These stringy considerations provide a nice verification of our findings above that the extremal black string will always nucleate a bubble.  In the extremal limit, the circle shrinks to zero at the degenerate horizon, so tachyon condensation takes over before extremality is reached and a bubble is nucleated.  Sufficiently far from extremality it becomes increasingly difficult to nucleate a bubble if the size of the extra dimension is appropriately large.  For the static bubble of minimum compactification radius ($R_s=4P$, $r_+=\sqrt{2}P$) for which all nucleated bubbles will expand out to infinity, we find that $\ell_s^2=8P^2$ which is inconsistent with our requirement that $\alpha'$ corrections can be ignored; we must choose $R_s^2\gg P^2$ which means we can ignore any ``catastrophic" bubbles.  Stated concisely, stringy analysis suggests that magnetic black strings will Hawking evaporate until they approach extremality sufficiently closely, at which point tachyon condensation will cause the circle to pinch off.  This destroys the horizon, leaving a bubble which settles down to the small static configuration considered above.  Similar results were found in \cite{Horowitz:2005vp} where black strings with F1 charge and black strings with F1 and NS5 charges also settle down to stable static bubbles.\footnote{We would like to thank an anonymous referee for bringing this paper to our attention.}  We would like to stress, however, that our current analysis makes use only of the  Einstein-Maxwell equations and does not require the properties of other string-inspired fields, such as the dilaton or various other gauge fields.

The ability to construct magnetically charged solitons from magnetic branes seems to be generic and directly carries over into 10D Einstein-Maxwell-Dilaton theory.  The dilaton has a non-trivial effect on the spacetime structure, for instance making the singularity for the charged black hole spacelike and as well as eliminating the problematic inner Cauchy horizon\cite{Garfinkle:1990qj}.  It was shown in \cite{Gregory:1994tw} that SUSY black p-branes are perturbatively stable, i.e. do not suffer the Gregory-Laflamme instability.  This result is insensitive to fermionic boundary conditions and holds for extremal black p-branes as well.  Further arguments were given in
favour of their thermodynamic stability \cite{Gregory:1994tw}; however
these were based on comparing the entropies of black p-branes to spherical black holes but these objects are charged under different rank gauge fields.    In much the same way as black p-branes with RR charge were considered in \cite{Horowitz:2005vp}, it would be interesting to revisit the question of thermodynamic stability of black p-branes with magnetic charge to see if they suffer a similar fate. 
\bigskip

We would like to thank the Natural Sciences and Engineering Research Council of Canada for providing the funding for this research.  We would also like to thank Keith Copsey for very valuable discussions.


\begin{thebibliography}{30}
\expandafter\ifx\csname natexlab\endcsname\relax\def\natexlab#1{#1}\fi
\expandafter\ifx\csname bibnamefont\endcsname\relax
  \def\bibnamefont#1{#1}\fi
\expandafter\ifx\csname bibfnamefont\endcsname\relax
  \def\bibfnamefont#1{#1}\fi
\expandafter\ifx\csname citenamefont\endcsname\relax
  \def\citenamefont#1{#1}\fi
\expandafter\ifx\csname url\endcsname\relax
  \def\url#1{\texttt{#1}}\fi
\expandafter\ifx\csname urlprefix\endcsname\relax\def\urlprefix{URL }\fi
\providecommand{\bibinfo}[2]{#2}
\providecommand{\eprint}[2][]{\url{#2}}

  
     \bibitem[{\citenamefont{Hawking et~al.}(1983)\citenamefont{Hawking and Page}}]{Hawking:1982dh}
\bibinfo{author}{\bibfnamefont{S.~W.} \bibnamefont{Hawking}}, \bibnamefont{and}
  \bibinfo{author}{\bibfnamefont{D.~N.} \bibnamefont{Page}},
  \bibinfo{journal}{Commun. Math. Phys.} \textbf{\bibinfo{volume}{87}},
  \bibinfo{pages}{577} (\bibinfo{year}{1983}).
  
      \bibitem[{\citenamefont{Gregory et~al.}(1993)\citenamefont{Gregory and Laflamme}}]{Gregory:1993vy}
\bibinfo{author}{\bibfnamefont{R.} \bibnamefont{Gregory}}, \bibnamefont{and}
  \bibinfo{author}{\bibfnamefont{R.} \bibnamefont{Laflamme}},
  \bibinfo{journal}{Phys. Rev. Lett.} \textbf{\bibinfo{volume}{70}},
  \bibinfo{pages}{2837} (\bibinfo{year}{1993}).
  
       \bibitem[{\citenamefont{Witten}(1998)\citenamefont{Witten}}]{Witten:1998wn}
\bibinfo{author}{\bibfnamefont{E.} \bibnamefont{Witten}},
  \bibinfo{journal}{Adv. Theor. Math. Phys.} \textbf{\bibinfo{volume}{2}},
  \bibinfo{pages}{505} (\bibinfo{year}{1998}).
  
       \bibitem[{\citenamefont{Cardoso et~al.}(1993)\citenamefont{Cardoso and Dias}}]{Cardoso:2006ks}
\bibinfo{author}{\bibfnamefont{V.} \bibnamefont{Cardoso}}, \bibnamefont{and}
  \bibinfo{author}{\bibfnamefont{O.~J.~C.} \bibnamefont{Dias}},
  \bibinfo{journal}{Phys. Rev. Lett.} \textbf{\bibinfo{volume}{96}},
  \bibinfo{pages}{181601} (\bibinfo{year}{2006}).
  
       \bibitem[{\citenamefont{Gubser et~al.}(2001)\citenamefont{Gubser and Mitra}}]{Gubser:2000mm}
\bibinfo{author}{\bibfnamefont{S.~S.} \bibnamefont{Gubser}}, \bibnamefont{and}
  \bibinfo{author}{\bibfnamefont{I.}~\bibnamefont{Mitra}},
  \bibinfo{eprint}{JHEP} \textbf{\bibinfo{volume}{0108}},
  \bibinfo{pages}{018} (\bibinfo{year}{2001}).
  
  \bibitem[{\citenamefont{Friess et~al.}(2005)\citenamefont{Friess, Gubser and Mitra}}]{Friess:2005zp}
\bibinfo{author}{\bibfnamefont{J.~J.} \bibnamefont{Friess}},
  \bibinfo{author}{\bibfnamefont{S.~S.}~\bibnamefont{Gubser}}, \bibnamefont{and}
  \bibinfo{author}{\bibfnamefont{I.} \bibnamefont{Mitra}},
  \bibinfo{journal}{Phys. Rev.} \textbf{\bibinfo{volume}{D72}},
  \bibinfo{pages}{104019} (\bibinfo{year}{2005}).
  
   \bibitem[{\citenamefont{Miyamoto}(2008)\citenamefont{Miyamoto}}]{Miyamoto:2007mh}
  \bibinfo{author}{\bibfnamefont{U.} \bibnamefont{Miyamoto}},
  \bibinfo{journal}{Phys. Lett. } \textbf{\bibinfo{volume}{B659}},
  \bibinfo{pages}{380} (\bibinfo{year}{2008}).
  
   \bibitem[{\citenamefont{Brill et~al.}(1991)\citenamefont{Brill and Horowitz}}]{Brill:1991qe}
\bibinfo{author}{\bibfnamefont{D.}~\bibnamefont{Brill}}, \bibnamefont{and}
  \bibinfo{author}{\bibfnamefont{G.~T.} \bibnamefont{Horowitz}},
  \bibinfo{journal}{Phys. Lett.} \textbf{\bibinfo{volume}{B262}},
  \bibinfo{pages}{437} (\bibinfo{year}{1991}).
  
    \bibitem[{\citenamefont{Witten}(1982)\citenamefont{Witten}}]{Witten:1981gj}
  \bibinfo{author}{\bibfnamefont{E.} \bibnamefont{Witten}},
  \bibinfo{journal}{Nucl. Phys.} \textbf{\bibinfo{volume}{B195}},
  \bibinfo{pages}{481} (\bibinfo{year}{1982}).
  
     \bibitem[{\citenamefont{Sarbach et~al.}(2004)\citenamefont{Sarbach and Lehner}}]{Sarbach:2004rm}
\bibinfo{author}{\bibfnamefont{O.}~\bibnamefont{Sarbach}}, \bibnamefont{and}
  \bibinfo{author}{\bibfnamefont{L.} \bibnamefont{Lehner}},
  \bibinfo{journal}{Phys. Rev.} \textbf{\bibinfo{volume}{D71}},
  \bibinfo{pages}{026002} (\bibinfo{year}{2005}).
  
    \bibitem[{\citenamefont{Sarbach et~al.}(2004)\citenamefont{Sarbach and Lehner}}]{Sarbach:2003dz}
\bibinfo{author}{\bibfnamefont{O.}~\bibnamefont{Sarbach}}, \bibnamefont{and}
  \bibinfo{author}{\bibfnamefont{L.} \bibnamefont{Lehner}},
  \bibinfo{journal}{Phys. Rev.} \textbf{\bibinfo{volume}{D69}},
  \bibinfo{pages}{021901} (\bibinfo{year}{2004}).
  
  \bibitem[{\citenamefont{Dine et~al.}(2009)\citenamefont{Dine, Shomer and Sun}}]{Dine:2006we}
\bibinfo{author}{\bibfnamefont{M.} \bibnamefont{Dine}},
  \bibinfo{author}{\bibfnamefont{A.}~\bibnamefont{Shomer}}, \bibnamefont{and}
  \bibinfo{author}{\bibfnamefont{Z.} \bibnamefont{Sun}},
  \bibinfo{journal}{JHEP} \textbf{\bibinfo{volume}{0612}},
  \bibinfo{pages}{013} (\bibinfo{year}{2006}).
  
     \bibitem[{\citenamefont{Kim et~al.}(2009)\citenamefont{Kim, Hornlund, Palmkvist and Virmani}}]{Kim:2010bf}
\bibinfo{author}{\bibfnamefont{S.~-S.} \bibnamefont{Kim}},
  \bibinfo{author}{\bibfnamefont{J.~L.}~\bibnamefont{Hornlund}},
  \bibinfo{author}{\bibfnamefont{J.}~\bibnamefont{Palmkvist}}, \bibnamefont{and}
  \bibinfo{author}{\bibfnamefont{A.} \bibnamefont{Virmani}},
  \bibinfo{journal}{JHEP} \textbf{\bibinfo{volume}{1008}},
  \bibinfo{pages}{072} (\bibinfo{year}{2010}).
    
      \bibitem[{\citenamefont{Mann et~al.}(2006)\citenamefont{Mann, Marolf and Virmani}}]{Mann:2006bd}
\bibinfo{author}{\bibfnamefont{R.~B.} \bibnamefont{Mann}},
  \bibinfo{author}{\bibfnamefont{D.}~\bibnamefont{Marolf}}, \bibnamefont{and}
  \bibinfo{author}{\bibfnamefont{A.} \bibnamefont{Virmani}},
  \bibinfo{journal}{Class. Quant. Grav.} \textbf{\bibinfo{volume}{23}},
  \bibinfo{pages}{6357} (\bibinfo{year}{2006}).

        \bibitem[{\citenamefont{Mann et~al.}(2006)\citenamefont{Mann and Marolf}}]{Mann:2005yr}
\bibinfo{author}{\bibfnamefont{R.~B.} \bibnamefont{Mann}}, \bibnamefont{and}
  \bibinfo{author}{\bibfnamefont{D.} \bibnamefont{Marolf}},
  \bibinfo{journal}{Class. Quant. Grav.} \textbf{\bibinfo{volume}{23}},
  \bibinfo{pages}{2927} (\bibinfo{year}{2006}).
  
      \bibitem[{\citenamefont{Rohm}(1984)\citenamefont{Rohm}}]{Rohm:1983aq}
  \bibinfo{author}{\bibfnamefont{R.}~\bibnamefont{Rohm}},
  \bibinfo{journal}{Nucl. Phys.} \textbf{\bibinfo{volume}{B237}},
  \bibinfo{pages}{553} (\bibinfo{year}{1984}).

     \bibitem[{\citenamefont{Adams et~al.}(2005)\citenamefont{Adams, Liu, McGreevy, Saltman and Silverstein}}]{Adams:2005rb}
  \bibinfo{author}{\bibfnamefont{A.}~\bibnamefont{Adams}}, 
  \bibinfo{author}{\bibfnamefont{X.}~\bibnamefont{Liu}}, 
  \bibinfo{author}{\bibfnamefont{J.}~\bibnamefont{McGreevy}}, 
  \bibinfo{author}{\bibfnamefont{A.}~\bibnamefont{Saltman}}, \bibnamefont{and}
  \bibinfo{author}{\bibfnamefont{E.} \bibnamefont{Silverstein}},
  \bibinfo{journal}{JHEP} \textbf{\bibinfo{volume}{0510}},
  \bibinfo{pages}{033} (\bibinfo{year}{2005}).
 
     \bibitem[{\citenamefont{Horowitz}(2005)\citenamefont{Horowitz}}]{Horowitz:2005vp}
  \bibinfo{author}{\bibfnamefont{G.}~\bibnamefont{Horowitz}},
  \bibinfo{journal}{JHEP} \textbf{\bibinfo{volume}{0508}},
  \bibinfo{pages}{091} (\bibinfo{year}{2005}).

  
        \bibitem[{\citenamefont{Garfinkle et~al.}(1991)\citenamefont{Garfinkle, Horowitz and Strominger}}]{Garfinkle:1990qj}
\bibinfo{author}{\bibfnamefont{D.} \bibnamefont{Garfinkle}},
  \bibinfo{author}{\bibfnamefont{G.~T.}~\bibnamefont{Horowitz}}, \bibnamefont{and}
  \bibinfo{author}{\bibfnamefont{A.} \bibnamefont{Strominger}},
  \bibinfo{journal}{Phys. Rev.} \textbf{\bibinfo{volume}{D43}},
  \bibinfo{pages}{3140} (\bibinfo{year}{1991}).
      
     \bibitem[{\citenamefont{Gregory et~al.}(1995)\citenamefont{Gregory and Laflamme}}]{Gregory:1994tw}
  \bibinfo{author}{\bibfnamefont{R.}~\bibnamefont{Gregory}}, \bibnamefont{and}
  \bibinfo{author}{\bibfnamefont{R.} \bibnamefont{Laflamme}},
  \bibinfo{journal}{Phys. Rev.} \textbf{\bibinfo{volume}{D51}},
  \bibinfo{pages}{305} (\bibinfo{year}{1995}).
  

  
\end{thebibliography}
\end{document}